\documentclass[epj]{svjour}
\usepackage{graphicx}

\begin{document}

\title{Gauge invariance in fractional  field theories}

\author{Richard Herrmann\inst{} 
\thanks{\emph{email address:} herrmann@gigahedron.com}%
}                     
\institute{GigaHedron, Farnweg 71, D-63225 Langen, Germany}
\date{Revised: {2008-06-06} / Generated: {\today}}
\abstract{
The principle of local gauge invariance is applied to fractional wave equations and 
the interaction term  is determined up to order $o(\bar{g})$ in the coupling constant $\bar{g}$.
As a first application, based on the Riemann-Liouville fractional derivative definition, 
the fractional Zeeman effect is used to reproduce the baryon spectrum accurately.
 The transformation
properties of the non relativistic fractional Schr\"odinger-equation under spatial rotations are
investigated and an internal fractional spin is deduced.
} 
\maketitle
\section{Introduction}
Historically the first example for a quantum field theory is quantum electro dynamics (QED), which successfully
 describes the interaction between
electrons and positrons with photons \cite{pauli},\cite{sakurai}. The interaction 
type of this quantum field theory between the electromagnetic
field and the electron Dirac field is determined by the principle of minimal gauge invariant coupling of the
electric charge. 

For more complex charge types, this principle may be extended to 
non abelian gauge theories or Yang-Mills field theories \cite{yang}.  
A typical example is quantum chromo dynamics (QCD), which describes a hadronic interaction by gauge invariant
coupling of
quark fields with gluon fields. 

Hence local gauge invariance seems to be a fundamental principle for a study of the dynamics of particles \cite{lee}.

In the following we will apply the concept of local gauge invariance to fractional fields and
will outline the basic structure of a quantum field theory based on fractional calculus.

The fractional calculus has fascinated mathematicians since the days of Leibniz \cite{f1}-\cite{riemann}. 
The interest in fractional relativistic quantum wave equations is a relatively new one. In 2000, Raspini
\cite{raspini},\cite{raspini1} has proposed 
a fractional Dirac equation of order $\alpha = 2/3$ and found the corresponding $\gamma_\alpha^\mu$-matrix
algebra to be related to generalized Clifford algebras. 
Z$\acute{a}$vada \cite{zavada} has generalized this approach and found, that relativistic covariant equations generated 
by taking the n-th root of the d'Alembert operator are fractional wave equations with an inherent
$SU(n)$ symmetry.

Since the only mechanism  currently used to introduce a symmetry like $SU(n)$
involves non abelian gauge fields as proposed
by Yang-Mills, the study of gauge invariant fractional wave equations is a new,  interesting alternative approach to
establish a $SU(n)$ symmetry.

Until now, only free fractional fields have been studied \cite{lim},\cite{gold}. 
In this paper, we will apply the principle of local gauge invariance to fractional wave equations and derive
the resulting interaction up to order $o(\bar{g})$ in the coupling constant $\bar{g}$. As an application, 
we will use the
fractional generalization of the classical Zeeman effect \cite{zeeman} for non relativistic 
 particles to reproduce the spectrum of
baryons accurately.    
\section{Notation}
The fractional calculus \cite{f3}-\cite{he08} provides a set of axioms and methods to extend the coordinate and corresponding
derivative definitions in a reasonable way from integer order n to arbitrary order $\alpha$:
\begin{equation}
\{ x^n, {\partial^n \over \partial x^n} \} 
\rightarrow
\{ x^\alpha, {\partial^\alpha \over \partial x^\alpha} \}
\end{equation}
The definition of the fractional order derivative is not unique,  several definitions 
e.g. the Riemann, Caputo, Weyl, Riesz, Feller,  Gr\"unwald  fractional 
derivative definition coexist \cite{caputo}-\cite{pod}.
To keep this paper as general as possible,   we do not apply a 
specific representation of the fractional derivative operator.

We will only assume, that an appropriate mapping on real numbers of coordinates $x$ and fractional coordinates $x^\alpha$ 
and
functions $f$ and fractional derivatives $g$  exists  and
that a Leibniz product rule is defined properly.

Therefore we use $x^\alpha$ as a short hand notation for e.g.
$sign(x) |x|^\alpha$ as demonstrated in \cite{he07} and $\partial^\alpha / \partial x^\alpha$ as a 
short hand notation for e.g. the fractional
left and right Riemann Liouville derivative ($D_{a+}^\alpha,D_{a-}^\alpha$) as demonstrated in \cite{baleanu}.
\begin{eqnarray} 
\label{riemann1}
(D_{a+}^\alpha f)(x) &=&  
\frac{1}{\Gamma(1 -\alpha)} \frac{\partial}{\partial x}  
     \int_a^x  d\xi \, (x-\xi)^{-\alpha} f(\xi)\\
\label{riemann2}
(D_{a-}^\alpha f)(x) &=&  
\frac{1}{\Gamma(1 -\alpha)} \frac{\partial}{\partial x}  
     \int_x^a  d\xi \, (\xi-x)^{-\alpha} f(\xi)
\end{eqnarray} 
Only in 
section \ref{szeeman}, 
where the fractional Zeeman effect is examined, we explicitely use the Riemann-Liouville fractional
derivative.

The Leibniz product rule is used in the following form \cite{f3},\cite{o1}:
\begin{equation}
\label{leibniz}
{\partial^\alpha \over \partial x^\alpha} (\phi \psi) = \sum_{k=0}^{\infty} 
\left(\begin{array}{c}
\alpha \\ k \\
\end{array} \right)
( {\partial^k \over \partial x^k} \phi ) 
( {\partial^{\alpha-k} \over \partial x^{\alpha-k}}\psi ) 
\end{equation}
where the fractional binomial is given by
\begin{equation}
\left(\begin{array}{c}
\alpha \\ k \\
\end{array} \right)
= {\Gamma(1+\alpha) \over \Gamma(1+k) \Gamma(1+\alpha-k)}
\end{equation}
and $\Gamma(z)$ is the Euler $\Gamma$-function.

Since until now, there exists no specific theory of fractional tensor calculus, in the following we propose
notations, which turn out to be useful. 

Especially the fractional derivative and in general  the quantities we will investigate 
are elements of a direct product space of the
discrete coordinate space on one hand, whose discrete coordinate indices are denoted by
$\mu,\nu,..= 0,..,3$ and on the other hand the continuous space of fractional derivatives, which are labeled with
$\alpha,\beta,... \in \cal{R}$. These are real numbers, which in the case $\alpha<0$ or $\beta<0$ are interpreted as a 
fractional integral of order $\alpha$ and $\beta$ respectively.

For the covariant fractional derivatives of order $\alpha$ we introduce the  short hand notations
\begin{equation}
{\partial^\alpha \over \partial (x^\mu)^\alpha} f = \partial_\mu^\alpha f = f^\alpha_{|\mu} 
\end{equation}
Einsteins summation convention for tensors e.g.
\begin{equation}
\partial_\mu = g_{\mu \nu}\partial^\nu = \sum_{\nu=0}^3 g_{\mu \nu}\partial^\nu 
\end{equation}
is extended to the continuous case for the fractional derivative coefficients
\begin{equation}
\label{cg}
\partial_\alpha = g_{\alpha \beta}\partial^\beta = \int_{-\infty}^{+\infty}  \!\! d\beta \hat{g}(\alpha,\beta)\partial^\beta 
\end{equation}
where $\hat{g}(\alpha, \beta)$ is an appropriately chosen  function of the two variables $\alpha,\beta$. 

Hence we define the generalized metric tensor 
\begin{equation}
\eta^{\mu\nu}_{\alpha \beta} = \textrm{diag}\{-1,1,1,1\}\delta(\alpha-\beta)
\end{equation}
where $\delta(\alpha-\beta)$ is the Dirac delta function for raising and lowering of indices.

Therefore the fractional derivative objects are formally described in a  manner similar to the case of spinor objects.
A quantity like e.g. $F^\alpha_{\mu \nu}$ behaves as a tensor of rank 2 in coordinate space and as a 
vector in the sense of (\ref{cg}) in fractional 
derivative space.    

With these definitions, the contravariant fractional derivative follows as:
\begin{equation}
 \partial_\alpha^\mu f = f^{|\mu}_{\alpha} = 
\eta^{\mu\nu}_{\alpha \beta}\partial_\nu^\beta f
\end{equation}
Calculations are performed in natural units $\hbar = c = 1$. 

Finally, we have to resolve an ambiguity, which is specific to fractional tensor calculus. For $\alpha=0$ the 
use of the fractional derivative introduces an index and therefore
\begin{equation}
\label{vio}
 \partial_\mu^0 f = f
\end{equation}
obviously violates the principle of covariance, since the  tensor  rank differs for left and right hand side of (\ref{vio}).
Consequently  we introduce the Kronecker delta $\delta_0^\mu$: 
\begin{equation}
\label{vio2}
\delta_0^\mu \partial_\mu^0 f =\delta^0_\mu \partial^\mu_0 f = f 
\end{equation}
and perform a pseudo-summation, which is reduced to a single term. 
We introduce the abbreviations
\begin{eqnarray}
\label{vioxx}
\delta_0^\mu \partial_\mu^0  &=& \delta^{(\mu)} \\
\delta^0_\mu \partial^\mu_0  &=& \delta_{(\mu)} 
\end{eqnarray}
and call ${(\mu)}$ a pseudo index, where summation is performed, but the tensor properties in fractional coordinate 
space are not affected.

The Kronecker delta may be generalised to
\begin{equation}
\label{vio4}
{\delta_{(\mu)}}^\tau  \partial_\tau^\alpha f = \partial_{(\mu)}^\alpha f 
\end{equation}
where $\partial_{(\mu)}^\alpha f $ transforms like a scalar in fractional coordinate space, but the derivative
with respect to $\mu$ is performed. With this definition, the Kronecker delta will be used like  
\begin{equation}
{\delta_{(\mu)}}^{\tau\sigma}  \partial_\tau^\alpha \partial_\sigma^{-\beta} f =   
{\delta_{(\mu)}}^\tau  \partial_\tau^{\alpha-\beta}  f = 
\partial_{(\mu)}^{\alpha -\beta} f   
\end{equation}
Hence introducing
the Kronecker delta  extends manifest covariant tensor calculus in a reasonable way to fractional derivative 
tensors. 
\section{Generalized Euler-Lagrange equations for fractional fields}
Euler-Lagrange equations for the Riemann-Liouville partial fractional derivative have been derived by 
Baleanu and Muslih \cite{baleanu} for classical fields, described by a Lagrange density $ {\cal{L}}$ of type
 ${\cal{L}} = {\cal{L}}(\phi, \phi_{|\mu}^\alpha)$. 

For our purpose, an extended version of the Lagrange density for a field $\phi$
is needed, which is given by  
\begin{equation}
\label{ltype}
{\cal{L}} = {\cal{L}}
(\phi, \phi_{|\mu}^\alpha, \phi_{|\mu}^{-\alpha},\phi^{-\alpha+1}_{|\mu},...,\phi^{-\alpha+k}_{|\mu})
\end{equation}
Following \cite{baleanu},\cite{baleanu1} variation of the covariant action $S = \int d^4 x{\cal{L}}$ 
yields the corresponding Euler-Lagrange equations 
\begin{equation}
\label{lsystem}
{ \partial {\cal{L}} \over  \partial \phi}
 - \partial_\mu^{\alpha} {\partial {\cal{L}} \over  \partial(\phi^{\alpha}_{|\mu})}
- \sum_{n=0}^k (-1)^n \partial_\mu^{-\alpha +n}
{ \partial {\cal{L}} \over  \partial(\phi^{-\alpha+n}_{|\mu})} = 0  
\end{equation}
Since the Lagrange density is non local, special care should be taken handling
 the corresponding Hamiltonian\cite{baleanu}. 
\section{Free fractional fields}
\subsection{The free matter field}
As an example for a  Lorentz-invariant, relativistic  free fractional field equation we present the fractional
extension of the Dirac equation.
 
Setting 
\begin{equation}
\label{alpha}
\alpha = 2/n 
\end{equation}
with $n = 2,3,4, ..$, the fractional Dirac equation results from the n-fold factorization 
of the relativistic energy-momentum relation 
\begin{equation}
\hat{E}^2 = \hat{p}^2 + m^2
\end{equation}
and is given by \cite{raspini},\cite{raspini1}:
\begin{equation}
(i \gamma_\alpha^\mu \partial_\mu^\alpha + {\bf{1}}_\alpha m^\alpha)\Psi = 0
\end{equation}
The $\gamma_\alpha^\mu$-matrices are built from triads of traceless, unitary $n \times n$ matrices, which
span a subspace of $SU(n)$ and ${\bf{1}}_\alpha$ is the corresponding  unit matrix. They obey an extended Clifford algebra  \cite{zavada}:
\begin{equation}
\sum_{\{\pi\}} \prod_{i=1}^{n}\gamma_\alpha^{\mu_i} = n! \delta^{\mu_1 \mu_2...\mu_n}
\end{equation}
where $\{\pi\}$ denotes all permutations of $\gamma_\alpha^{\mu_i}$ and $\delta^{\mu_1 \mu_2...\mu_n}$ denotes the
 Kronecker delta.

The corresponding Lagrange density is given by \cite{baleanu}:
\begin{equation}
{\cal{L}}_D = \bar{\Psi}( i \gamma_\alpha^{\mu} \partial_\mu^\alpha + {\bf{1}}_\alpha m^\alpha) \Psi 
\end{equation}
\subsection{The free gauge field}
As an example for an abelian fractional gauge field we present the fractional extension of the
Maxwell equations. 
We define the fractional field strength tensor of rank 2 in coordinate space and rank 1 in fractional derivative space by
\begin{equation}
\label{fem}
F_{\mu\nu}^\alpha  = \partial_\nu^\alpha A_\mu - \partial_\mu^\alpha A_\nu
\end{equation}
The Lagrange density for the fractional electro-magnetic field ${\cal{L}}_{EM}$  is given by 
\begin{equation}
\label{lem}
{\cal{L}}_{EM} = -{1 \over 4} F_{\mu\nu}^\alpha F_{\alpha}^{\mu \nu}
\end{equation}
Variation  of ${\cal{L}}_{EM}$  with respect to the 
fractional potential $A_\mu$ yields  the fractional inhomogenous  Maxwell equations
\begin{equation}
\label{inhomogenous}
\partial_\mu^\alpha F_{\alpha}^{ \mu \nu} = 0
\end{equation}
The potential $A_\mu$ is not uniquely determined. A change of the potential of type  
\begin{equation}
 A^{'}_\mu  = A_\mu + \partial_\mu^\alpha \phi = A_\mu + \phi^\alpha_{|\mu}   
\end{equation}
leaves the field strength tensor unchanged and is therefore called a gauge transformation.

\section{Gauge invariance in first order of the coupling constant $\bar{g}$ }
In the previous section we have presented an example for a fractional free field, the relativistic fractional 
Dirac field and for a fractional gauge field, the fractional electromagnetic field.
 
We now postulate, that the principle of local gauge invariance is valid for fractional
wave equations, too.  

The requirement of invariance of the fractional Lagrange density for the  Dirac field under
 local gauge transformations should
uniquely determine the type of the interaction with a fractional gauge field.  

We therefore will investigate the transformation properties of the fractional analogue of 
the free QED-Lagran\-gian ${\cal{L}}_{FQED}^{free}$:
\begin{equation}
\label{fqedfree}
{\cal{L}}_{FQED}^{free} = \bar{\Psi}( i \gamma_\alpha^{\mu} \partial_\mu^\alpha + {\bf{1}}_\alpha m^\alpha) \Psi 
 -{1 \over 4} F_{\mu\nu}^\beta F_{\beta}^{ \mu \nu}
\end{equation}
where $\alpha=2/n$ and $\beta$ both denote  derivatives of fractional order.
Special cases are: For $\alpha=1$ a fractional electric field
couples to the standard Dirac field, for $\beta=1$ a standard electric field couples to a fractional Dirac field. For
$\alpha=\beta$ both fractional fields are of similar type. For $\alpha=\beta=1$ the Lagrange density (\ref{fqedfree})
reduces to
the QED Lagrangian. 

Under local gauge transformations the following transformation properties hold:
\begin{eqnarray}
\bar{\Psi}^{'}   &=& \bar{\Psi} e^{-i \bar{g}\phi(x)} \\ 
\Psi^{'}         &=& e^{i \bar{g}\phi(x)} \Psi \\ 
A_{\mu}^{'}      &=& A_{\mu} + \phi_{|\mu}^{\beta} 
\end{eqnarray}
The transformation properties of the fractional derivative operator $\partial^\alpha_\mu$ may be deduced from: 
\begin{eqnarray}
\bar{\Psi}^{'} \gamma_\alpha^{\mu} \partial_\mu^\alpha \Psi^{'}    
&=& \bar{\Psi}^{'} \gamma_\alpha^{\mu}
 \partial_\mu^\alpha e^{i \bar{g}\phi} \Psi \\
&=& \bar{\Psi}^{'} \gamma_\alpha^{\mu}
(e^{i \bar{g}\phi} \partial_\mu^\alpha  + [ \partial_\mu^\alpha ,e^{i \bar{g}\phi} ])\Psi \\
&=& \bar{\Psi} \gamma_\alpha^{\mu}
(\partial_\mu^\alpha  +  e^{-i \bar{g}\phi}  [ \partial_\mu^\alpha ,e^{i \bar{g}\phi} ])\Psi 
\end{eqnarray}
The commutator reflects the non locality for the fractional derivative. Since there is no simple fractional analogue
to the chain rule, which is known for the  standard derivative, this commutator cannot be evaluated directly, without
loss of clarity.  

Therefore we will restrict our derivation to 
gauge invariance up to first order in the coupling constant $\bar{g}$ only, which corresponds to an
infinitesimal gauge transformation.
 
We obtain:
\begin{eqnarray}
 e^{-i \bar{g}\phi}  [ \partial_\mu^\alpha ,e^{i \bar{g}\phi} ]  
&=& (1 -i \bar{g}\phi) [ \partial_\mu^\alpha ,1 + i \bar{g}\phi ] \\
&=& (1 -i \bar{g}\phi) [ \partial_\mu^\alpha ,i \bar{g}\phi ]   \\
&=& i \bar{g} [ \partial_\mu^\alpha ,\phi ] \quad\quad \quad \quad\quad \quad  + o(\bar{g}^2)      
\end{eqnarray}
We define the fractional charge     connection operator $\hat{\Gamma}^{\alpha \beta}_\mu$, which 
in a fractional charge tangent space
is the 
analogue  to the fractional Christoffel symbols in a fractional 
coordinate tangent space or the fractional 
Fock-Ivanenko coefficients in a fractional spinor tangent space: 
\begin{eqnarray}
\label{g1}
\hat{\Gamma}^{\alpha \beta}_\mu(\xi_\mu) &=& {\delta_{(\mu)}}^{\sigma \tau} 
[ \partial_\sigma^\alpha ,( \partial_\tau^{-\beta} \xi_{\mu}) ]\\
\label{g2}
 &=& {\delta_{(\mu)}}^{\sigma \tau} [ \partial_\sigma^\alpha ,\xi^{-\beta}_{\mu|\tau} ]
\end{eqnarray}
with the Kronecker delta ${\delta_{(\mu)}}^{\sigma \tau}$, where $(\mu)$ indicates  that a summation over $\mu$
is performed, which reduces the sum (\ref{g1}) to a single term (\ref{g2}), but on the other hand 
$(\mu)$ should be ignored, when the transformation properties in fractional coordinate space are determined,
therefore the
vector character of $\hat{\Gamma}^{\alpha \beta}_\mu(\xi_\mu)$ is conserved in the fractional coordinate space.

This operator is linear
\begin{equation}
\hat{\Gamma}^{\alpha \beta}_\mu(\xi_\mu+ \zeta_\mu) = \hat{\Gamma}^{\alpha \beta}_\mu(\xi_\mu) + \hat{\Gamma}^{\alpha \beta}_\mu(\zeta_\mu)
\end{equation}
It should be emphasized, that $\hat{\Gamma}^{\alpha \beta}_\mu(\xi_\mu)$ is an operator and therefore does not transform as a 
simple c-number.
But up to first order in $\bar{g}$ the relation
\begin{equation}
\label{gam}
\bar{\Psi}^{'} \gamma_\alpha^{\mu} i \bar{g} \hat{\Gamma}^{\alpha \beta'}_\mu(\xi_\mu) \Psi^{'} = 
\bar{\Psi} \gamma_\alpha^{\mu} i \bar{g} \hat{\Gamma}^{\alpha \beta}_\mu(\xi_\mu) \Psi \quad +o(\bar{g}^2) 
\end{equation}
holds.
Hence the fractional derivative operator $\partial_\mu^\alpha$ transforms as
\begin{equation}
\bar{\Psi}^{'} \gamma_\alpha^{\mu} \partial_\mu^\alpha \Psi^{'} =    
\bar{\Psi} \gamma_\alpha^{\mu}(\partial_\mu^\alpha +  i \bar{g} \hat{\Gamma}^{\alpha \beta}_\mu(\phi^\beta_{|\mu})) \Psi \quad +o(\bar{g}^2) 
\end{equation}
Therefore we define the covariant fractional derivative $D_\mu^\alpha$ as 
\begin{equation}
D_\mu^\alpha =  \partial_\mu^\alpha -  i \bar{g} \hat{\Gamma}^{\alpha \beta}_\mu(A_\mu) 
\end{equation}
It transforms as
\begin{equation}
D_\mu^{\alpha'}  =
  \partial_\mu^\alpha 
+  i \bar{g} \hat{\Gamma}^{\alpha \beta}_\mu(\phi^\beta_{|\mu})
-  i \bar{g} \hat{\Gamma}^{\alpha \beta'}_\mu(A_\mu^{'})
\end{equation}
With the covariant fractional derivative $D_\mu^\alpha$ it follows from (\ref{gam}) and from the linearity of the 
fractional charge     connection operator $\hat{\Gamma}^{\alpha \beta}_\mu$, that 
 the Lagrange density ${\cal{L}}_{FQED}$ 
for the fractional extension of QED
 \begin{equation}
\label{ldirac}
{\cal{L}}_{FQED} = \bar{\Psi}( i \gamma_\alpha^{\mu} D_\mu^{\alpha} + {\bf{1}}_\alpha m^\alpha) \Psi  
-{1 \over 4} F_{\mu\nu}^\beta F_{\beta}^{\mu \nu}
\end{equation}
is invariant under local gauge transformations up to first order in the coupling constant $\bar{g}$.

Variation of ${\cal{L}}_{FQED}$ with respect to $\bar{\Psi}$ leads to the Dirac equation for the fractional Dirac field 
$\Psi$: 
\begin{equation}
\label{fdirac}
( i \gamma_\alpha^{\mu} \partial_\mu^{\alpha} + {\bf{1}}_\alpha m^\alpha) \Psi   = -\bar{g} \gamma_\alpha^\mu \hat{\Gamma}^{\alpha \beta}_\mu(A_\mu) \Psi 
\end{equation}
Variation of ${\cal{L}}_{FQED}$ with respect to the four potential $A_\mu$ leads to the fractional Maxwell equations
\begin{eqnarray}
\label{fmaxwell}
\partial_\mu^{\beta} F_{\beta}^{\mu \nu} 
&=& -\bar{g} \bar{\Psi} \gamma_\alpha^{\nu} ({\delta_{(\nu)}}^{\sigma \tau}
\partial_\tau^{-\beta} \partial_\sigma^{\alpha})  \Psi\\
& =& -\bar{g} \bar{\Psi}\gamma_\alpha^\nu  
({\delta_{(\nu)}}^\sigma \partial_\sigma^{\alpha-\beta}) 
 \Psi\\
\label{fmaxwell2}
    & =& -\bar{g} \bar{\Psi}\gamma_\alpha^\nu  \partial_{(\nu)}^{\alpha-\beta} \Psi
\end{eqnarray}
with ${\delta_{(\nu)}}^{\sigma\tau}$  is the Kronecker delta,
where $(\nu)$ indicates, that the term 
$({\delta_{(\nu)}}^{\sigma \tau}\partial_\tau^{-\beta} \partial_\sigma^{\alpha})  $ transforms like a scalar in fractional coordinate space 
but summation is performed in (\ref{fmaxwell}), which yields 
a single term $\partial_{(\nu)}^{\alpha-\beta}$ in (\ref{fmaxwell2}), which for $\alpha=\beta$ becomes $1$, while for
$\alpha > \beta$ fractional differentiation and for $\alpha < \beta$ fractional integration is performed. 

The term on the right side of (\ref{fmaxwell}) is the variation of 
\begin{equation}
{\cal{L}}_\Gamma = \bar{g} \bar{\Psi} \gamma_\alpha^{\mu} \hat{\Gamma}^{\alpha \beta}_\mu(A_\mu) \Psi 
\end{equation}
with respect to the four potential $A_\mu$ 
 and may be derived using the Leibniz product rule for fractional derivatives (\ref{leibniz}):
\begin{eqnarray}
{\cal{L}}_\Gamma &=& 
\bar{g} \bar{\Psi} \gamma_\alpha^{\mu}    \hat{\Gamma}^{\alpha \beta}_\mu(A_\mu) \Psi \\
&=& \bar{g} \bar{\Psi} \gamma_\alpha^{\mu} ({\delta_{(\mu)}}^{\sigma \tau} 
    [\partial_\sigma^\alpha , A^{-\beta}_{\mu|\tau}]) \Psi \nonumber \\
&=& \bar{g} \bar{\Psi} \gamma_\alpha^{\mu} ( {\delta_{(\mu)}}^{\sigma \tau}
   ( \partial_\sigma^\alpha A^{-\beta}_{\mu|\tau} - A^{-\beta}_{\mu|\tau} \partial_\sigma^\alpha))  \Psi \nonumber\\
&=& \bar{g} \bar{\Psi} \gamma_\alpha^{\mu}
({\delta_{(\mu)}}^{\sigma \tau}
(
\sum_{k=0}^{\infty}
\left(\begin{array}{c}
\alpha \\ k \\
\end{array} \right)
A^{-\beta+k}_{\mu|\tau} \partial^{\alpha - k}_\sigma 
 - A^{-\beta}_{\mu|\tau} \partial^\alpha_\sigma 
)
)
\Psi \nonumber\\ 
\label{lfinal}
&=& \bar{g} \bar{\Psi} \gamma_\alpha^{\mu}
({\delta_{(\mu)}}^{\sigma \tau}
\sum_{k=1}^{\infty}
\left(\begin{array}{c}
\alpha \\ k \\
\end{array} \right)
A^{-\beta+k}_{\mu|\tau} \partial^{\alpha - k}_\sigma 
)
\Psi 
\end{eqnarray}
Therefore ${\cal{L}}_\Gamma $ is a function of type (\ref{ltype}). 
Applying the corresponding Euler-Lagrange equations (\ref{lsystem}) to (\ref{lfinal}) yields
\begin{eqnarray}
 & & 
\sum_{k=1}^\infty (-1)^k \partial_\tau^{-\beta +k}
{ \partial {\cal{L}}_\Gamma \over  \partial(A^{-\beta+k}_{\mu|\tau})} 
 = \\  
&=& \bar{g} \bar{\Psi} \gamma_\alpha^{\mu} 
({\delta_{(\mu)}}^{\sigma \tau}
\sum_{k=1}^{\infty}
\left(\begin{array}{c}
\alpha \\ k \\
\end{array} \right)
(-1)^k \partial_\tau^{-\beta +k} \partial^{\alpha - k}_\sigma   
)
\Psi \\
&=& \bar{g} \bar{\Psi} \gamma_\alpha^{\mu}
({\delta_{(\mu)}}^{\sigma \tau}
\partial_\tau^{-\beta} \partial_\sigma^{\alpha}
\sum_{k=1}^{\infty}
\left(\begin{array}{c}
\alpha \\ k \\
\end{array} \right)
(-1)^k 
) 
\Psi \\
&=& -\bar{g} \bar{\Psi} \gamma_\alpha^{\mu} ({\delta_{(\mu)}}^{\sigma \tau}
\partial_\tau^{-\beta} \partial_\sigma^{\alpha}) 
\Psi 
\end{eqnarray}
Hence the coupling term in (\ref{fmaxwell}) is derived.

The non linear field equations (\ref{fdirac}) and (\ref{fmaxwell}) completely determine the interaction
of the fractional Dirac field with the fractional gauge field up to first order in $o(\bar{g})$. 
Therefore we have proven,
that the principle of local gauge invariance still is effective for fractional fields. 

Furthermore, besides the mechanism proposed by Yang-Mills, now there exists an alternative approach to implement an
inherent $SU(n)$-symmetry. While in a Yang-Mills theory, the $SU(n)$-symmetry is explicitely introduced via a non 
abelian gauge field, now 
the same symmetry is introduced implicitly via the fractional Dirac equation leaving the symmetry of the fractional 
gauge field abelian. Therefore equations (\ref{ldirac}),(\ref{fdirac}) and (\ref{fmaxwell}) serve as an
alternative formulation of a gauge field theory of e.g. for $\alpha=\beta=2/3$ the strong interaction of hadrons.

For $\alpha=1, \beta=1$ these equations reduce to the standard QED equations. 
For $\alpha \neq 1,\beta \neq 1$ the major
extensions are: the electric field $A_\mu$ extends from a c-number to a non local operator, which we 
called the fractional charge     connection operator. The
$\gamma_\alpha^\mu$-matrices obey an extended Clifford algebra, which reflects the transition from $SU(2)$ to $SU(n>2)$.
This makes the fractional QED an interesting candidate to describe properties of particles, which obey
an inherent $SU(n)$ symmetry. 

To help to get this idea accepted   we investigate in the following section 
the fractional extension of the classical Zeeman effect, 
which will be applied to a description of the baryon mass spectrum.
  
\section{A first application - the fractional Riemann Liouville Zeeman effect}
\label{szeeman}
In the previous section, we have applied a local gauge transformation to a free fractional Dirac field and 
obtained a fractional interaction with the fractional gauge field. Of course, this method is neither
restricted to spinor fields nor does it rely on Lorentz-covariant wave equations, 
but may be applied to other free fields as well.

In this section, we will present a simple example to illustrate  the validity of the proposed mechanism to generate an 
interaction for fractional  fields. 
We will investigate the fractional analogue of the classical Zeeman effect, which describes the level splitting of a
charged particle in an external constant magnetic field.

For that purpose,  we will first derive the fractional non-relativistic
Schr\"o\-din\-ger equation including the interaction term. 
\subsection{The fractional Schr\"odinger equation minimally coupled to an external fractional gauge field}
The Lagrange density ${\cal{L}}_{FS}^{free}$ for the free fractional  Schr\"o\-dinger field is given by
\begin{equation}
\label{sfree}
{\cal{L}}_{FS}^{free} = - {1 \over 2 m^{2 \alpha-1}}  
(\partial^i_{\alpha} \Psi)^*
(\partial_i^{\alpha} \Psi)
+i \Psi^* \partial_t \Psi
\end{equation}
Local gauge invariance up to first order in $\bar{g}$ is achieved by a replacement of the
fractional derivative by
\begin{eqnarray}
D_\mu^\alpha &=& \{D_t, D_i^\alpha\}
  = \{ \partial_t + i \bar{g} V,  \partial_i^\alpha -  i \bar{g} \hat{\Gamma}^{\alpha \beta}_i(A_i)\} \\ 
 &=& \{D_t, \vec{D}^\alpha\}
  = \{ \partial_t + i \bar{g} V,  \nabla^\alpha -  i \bar{g} \hat{\vec{\Gamma}}^{\alpha \beta} \} 
\end{eqnarray}
Therefore the free Lagrange density (\ref{sfree}) is extended to
\begin{equation}
{\cal{L}}_{FS} = - {1 \over 2 m^{2 \alpha-1}}  
(D^i_{\alpha} \Psi)^*
(D_i^{\alpha} \Psi)
+i \Psi^* (\partial_t + i \bar{g} V)\Psi
\end{equation}
Variation of ${\cal{L}}_{FS}$ with respect to $\Psi^*$ yields the fractional Schr\"odinger 
equation including an interaction with an external gauge field:
\begin{equation}
\label{sgl0}
 - {1 \over 2 m^{2 \alpha-1}}  
((\partial_i^\alpha -  i \bar{g} \hat{\Omega}^{\alpha \beta}_i) D^i_{\alpha} -i  (\partial_t + i \bar{g} V))\Psi=0
\end{equation}
or
\begin{equation}
- {1 \over 2 m^{2 \alpha-1}}  
(\partial_i^\alpha \partial^i_\alpha -  i \bar{g}( \hat{\Omega}^{\alpha \beta}_i \partial^i_\alpha 
+ \partial^i_\alpha \, \hat{\Gamma}^{\alpha \beta}_i)) \Psi = i  (\partial_t + i \bar{g} V)\Psi  
\end{equation}
or in vector notation
\begin{equation}
\label{sgl}
- {1 \over 2 m^{2 \alpha-1}}  
(\Delta^\alpha_\alpha -  i \bar{g}( \hat{\vec{\Omega}}^{\alpha \beta} \nabla_\alpha 
+ \nabla_\alpha \, \hat{\vec{\Gamma}}^{\alpha \beta})) \Psi = i  (\partial_t + i \bar{g} V)\Psi  
\end{equation}
where $\Delta^\alpha_\alpha=\nabla^\alpha \nabla_\alpha$ and $\hat{\Omega}^{\alpha \beta}_i$
\begin{equation}
\label{omega}
 \hat{\Omega}^{\alpha \beta}_i(A_i) =  
-{\delta_{(i)}}^{\sigma \tau}
\sum_{k=1}^{\infty}
\left(\begin{array}{c}
\alpha \\ k \\
\end{array} \right)
(-1)^k \partial_\sigma^{\alpha -k} A^{-\beta + k}_{i|\tau}   
\end{equation}
results from the variation of $\hat{\Gamma}^{\alpha \beta}_i(A_i)\Psi^*$.

With (\ref{sgl0})-(\ref{sgl}) the fractional Schr\"odinger equation including the interaction with an external
electro-magnetic field is derived, which is invariant under infinitesimal gauge transformations.
\subsection{The fractional Riemann-Liouville Zeeman effect}
We will now solve the derived fractional Sch\"odinger equation for a specific external field, which is the fractional analogue for a 
constant external magnetic field. In order to simplify the procedure, we will consider the special case
of similar fractional order of the fractional matter field and  
the fractional  electro-magnetic
field and set $\alpha=\beta$.

The  fractional charge     connection operator is then given according to (\ref{lfinal}) by
\begin{equation}
\label{sum}
\hat{\Gamma}^{\alpha \alpha}_i = 
\sum_{k=1}^{\infty}
\left(\begin{array}{c}
\alpha \\ k \\
\end{array} \right)
(\partial_i^{-\alpha+k} A_i)  \partial_i^{\alpha - k}
\end{equation}
and according (\ref{omega})
\begin{equation}
\label{OmegaSimply}
\hat{\Omega}^{\alpha \alpha}_i = 
-\sum_{k=1}^{\infty}
\left(\begin{array}{c}
\alpha \\ k \\
\end{array} \right)
(-1)^k
\partial_i^{\alpha - k}(\partial_i^{-\alpha+k} A_i)  
\end{equation}
In order to evaluate these series, we apply the Riemann-Liouville fractional derivative (\ref{riemann1}),(\ref{riemann2})
with $a=0$. 
\begin{equation}
\label{xxx}
\partial^\alpha x^\nu = {\Gamma(1+\nu) \over \Gamma(1 + \nu-\alpha)} x^{\nu-\alpha},\quad \nu-\alpha \ge -1 
\end{equation}
We now introduce the external field $\vec{A}$:
\begin{equation}
\label{aconst}
\vec{A}  = \{ - {1 \over 2} B x^{1-\alpha} y^{2\alpha-1} z^{\alpha-1},{1 \over 2} B x^{2\alpha-1} y^{1-\alpha}z^{\alpha-1},0 \}  
\end{equation}
which for $\alpha=1$  reduces to $ \{-\frac{1}{2}B y,\frac{1}{2}B x ,0 \} $ and is therefore the fractional analogue
for a constant magnetic field B. The fractional magnetic field $\vec{B}^\alpha$ is given by:
\begin{eqnarray}
\label{bconst}
\vec{B}^\alpha &=& \{B_x,B_y,B_z\} = \nabla^\alpha \times \vec{A}\\
        &=&  \{ 0,0, \frac{B}{2} \frac{\Gamma(2 \alpha)}{\Gamma(\alpha)} 
                     \left( x^{1-\alpha}y^{\alpha-1}+x^{\alpha-1}y^{1-\alpha} \right)z^{\alpha-1} \} \nonumber \\
        &=&  \{ 0,0, \frac{B}{2} \frac{\Gamma(2 \alpha)}{\Gamma(\alpha)} 
                     \left( (\frac{y}{x})^{\alpha-1}+ (\frac{x}{y})^{\alpha-1} \right)z^{\alpha-1} \} 
\end{eqnarray}
Since $1/\Gamma(0) = 0$, the following conditions hold for the z-component of the fractional magnetic field:
\begin{equation}
\label{zero}
\partial_x^\alpha \partial_y^\alpha B_z =\partial_y^\alpha \partial_x^\alpha B_z =\partial_z^\alpha B_z = 0
\end{equation}
Therefore, within the specific context of the Riemann-Liouville fractional
derivative definition (\ref{riemann1}),(\ref{riemann2}),  
the fractional magnetic field (\ref{bconst}) indeed is  the fractional extension of a constant electro-magnetic field.

For the fractional derivatives of $\vec{A}$ we obtain
\begin{equation}
\partial^{-\alpha+1} \vec{A} = \Gamma(2-\alpha) \{ - {1 \over 2} B  y^{2\alpha-1}z^{\alpha-1} ,{1 \over 2} B x^{2\alpha-1}z^{\alpha-1} ,0 \}
\end{equation}
and consecutively
\begin{equation}
\partial^k (\partial^{-\alpha+1} \vec{A}) = 0 
\end{equation}
As a consequence, the infinite series in (\ref{sum}) and (\ref{OmegaSimply}) each reduce to a single term.
\begin{eqnarray}
\hat{\vec{\Gamma}}^{\alpha \alpha} &=& 
\alpha\Gamma(2-\alpha){B \over 2}z^{\alpha-1}   \{ - y^{2\alpha-1} \partial_x^{\alpha-1},
 x^{2\alpha-1} \partial_y^{\alpha-1},
0 \}  \nonumber \\
&=& \hat{\vec{\Omega}}^{\alpha \alpha}  
\end{eqnarray}
We therefore obtain 
\begin{equation}
\nabla_\alpha \hat{\vec{\Gamma}}^{\alpha \alpha}=  \hat{\vec{\Omega}}^{\alpha \alpha} \nabla_\alpha 
\end{equation}
We introduce the fractional analogue of the z-component of the angular momentum operator $\hat{L}_z(\alpha)$
\begin{equation}
\hat{L}_z(\alpha) = i(y^\alpha \partial^\alpha_x - x^\alpha \partial^\alpha_y) 
\end{equation}
It follows:
\begin{equation}
 \nabla_\alpha \hat{\vec{\Gamma}}^{\alpha \alpha}  = \hat{\vec{\Omega}}^{\alpha \alpha} \nabla_\alpha = 
 i  \alpha\Gamma(2-\alpha) {B \over 2}  z^{\alpha-1}\hat{L}_z(2 \alpha-1)
\end{equation}
The fractional Schr\"odinger equation (\ref{sgl}) reduces to
\begin{equation}
 - {1 \over 2 m^{2 \alpha-1}}  
(\Delta^\alpha_\alpha  +  \bar{g}\alpha\Gamma(2-\alpha) B z^{\alpha-1}\hat{L}_z(2 \alpha-1) )\Psi = -i\partial_t \Psi
\end{equation}
With the product ansatz $\Psi = e^{-iEt}\psi$ a stationary Schr\"o\-di\-nger equation for
the energy spectrum results in:  
\begin{equation}
\label{zeeman}
H \psi =  - {1 \over 2 m^{2 \alpha-1}}  
(\Delta^\alpha_\alpha  +  \bar{g}\alpha\Gamma(2-\alpha) B z^{\alpha-1} \hat{L}_z(2\alpha-1)  )\psi = E \psi
\end{equation}
With (\ref{zeeman})  we have derived the non relativistic fractional Schr\"odinger equation 
for a spinless particle moving in a constant external 
fractional magnetic field, which is gauge invariant up to first order in $\hat{g}$. We want to emphasize, that
for a different choice of the fractional derivative definition a different Hamiltonian results, e.g. for 
the Caputo\cite{caputo} fractional derivative definition $z^{\alpha-1}$ has to be replaced by  $ 1$. 

\begin{figure}
\begin{center}
\includegraphics[width=80mm,height=40mm]{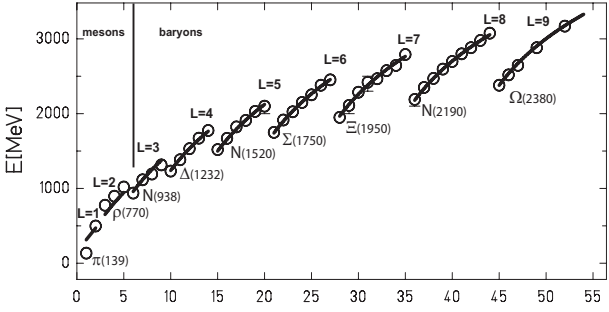}\\
\caption{\label{had}
comparison of the experimental hadron spectrum \cite{pdg} with a fit of 
the energy spectrum derived for the fractional Zeeman effect(\ref{zeemanlevels}). Experimental masses
are given as circles, theoretical values are given as lines for a given $L$. 
Band heads are labeled with their name and experimental mass from \cite{pdg}.
Note that the theoretical 
values for mesons $L=1,2$ are predicted values for baryon masses obtained with (\ref{zeemanlevels}).
} 
\end{center}
\end{figure}

\begin{table}
\begin{center}
\caption{\label{tone}
A comparison of the experimental baryon spectrum from \cite{pdg} with the mass formula (\ref{zeemanlevels}),
Listed are name, proposed quantum numbers $L$ and $M$,  experimental mass $E^{exp}[MeV]$, fitted
mass $E^{th}[MeV]$ and  error  $\Delta$E[$\%$]. For $L=1$ and $L=2$ the predicted baryon masses are compared
with experimental meson masses. * indicates the status of a particle. 
}
\begin{tabular}{lrrrrr}
\hline\noalign{\smallskip}
name& L & M & $E^{exp}[MeV]$ & $E^{th}[MeV]$ & $\Delta$E[$\%$] \\ 
\noalign{\smallskip}
\hline\noalign{\smallskip}
$\pi^0$ & 1 & 0 &135 & 313.90 & 132.57 \\ 
$K^0_s$ & 1 & 1 &498 & 470.45 & -5.46 \\ 
$\rho(770)$ & 2 & 0 &776 & 652.41 & -15.87 \\ 
$K^{*}(892)^0$ & 2 & 1 &896 & 808.96 & -9.71 \\ 
$\phi(1020)$ & 2 & 2 &1019 & 945.76 & -7.19 \\ 
\hline\noalign{\smallskip}
$N$ & 3 & 0 &938 & 959.39 & 2.28 \\ 
$\Lambda$ & 3 & 1 &1116 & 1115.94 & 0.02 \\ 
$\Sigma^0$ & 3 & 2 &1193 & 1252.74 & 5.04 \\ 
$\Xi^0$ & 3 & 3 &1315 & 1374.16 & 4.51 \\ 
$\Delta(1232)$ & 4 & 0 &1232 & 1240.53 & 0.69 \\ 
$\Sigma^0(1385)$ & 4 & 1 &1384 & 1397.08 & 0.97 \\ 
$\Xi(1530)$ & 4 & 2 &1532 & 1533.87 & 0.14 \\ 
$\Omega^{-}$ & 4 & 3 &1672 & 1655.29 & -1.00 \\ 
$\Lambda(1800)$ & 4 & 4 &1775 & 1764.44 & -0.60 \\ 
$\Lambda(1520)$ & 5 & 0 &1520 & 1500.10 & -1.28 \\ 
$\Lambda(1670)$ & 5 & 1 &1670 & 1656.65 & -0.80 \\ 
$\Xi(1820)$ & 5 & 2 &1823 & 1793.45 & -1.62 \\ 
$\Delta(1910)$ & 5 & 3 &1910 & 1914.87 & 0.25 \\ 
$\Sigma(2030)$ & 5 & 4 &2030 & 2024.01 & -0.29 \\ 
$\Lambda(2100)$ & 5 & 5 &2100 & 2123.14 & 1.10 \\ 
$\Sigma(1750)$ & 6 & 0 &1750 & 1741.42 & -0.49 \\ 
$\Sigma(1915)$ & 6 & 1 &1915 & 1897.97 & -0.89 \\ 
$\Xi(2030)$ & 6 & 2 &2025 & 2034.76 & 0.48 \\ 
$\Delta(2150)$(*) & 6 & 3 &2150 & 2156.18 & 0.29 \\ 
$\Omega(2250)$ & 6 & 4 &2252 & 2265.33 & 0.59 \\ 
$\Omega(2380)$(**) & 6 & 5 &2380 & 2364.46 & -0.65 \\ 
$\Sigma_c(2452)$ & 6 & 6 &2452 & 2455.27 & 0.13 \\ 
$\Xi(1950)$ & 7 & 0 &1950 & 1967.06 & 0.88 \\ 
$\Lambda(2110)$ & 7 & 1 &2110 & 2123.61 & 0.65 \\ 
$\Lambda_c$ & 7 & 2 &2286 & 2260.41 & -1.14 \\ 
$\Delta(2420)$ & 7 & 3 &2420 & 2381.83 & -1.58 \\ 
$\Xi^+_c(2467)$ & 7 & 4 &2467 & 2490.97 & 0.97 \\ 
$\Xi^{'+}_c(2575)$ & 7 & 5 &2576 & 2590.10 & 0.56 \\ 
$\Xi_c^0(2645)$ & 7 & 6 &2645 & 2680.92 & 1.35 \\ 
$\Xi_c^0(2790)$ & 7 & 7 &2791 & 2764.73 & -0.94 \\ 
$N(2190)$ & 8 & 0 &2190 & 2179.12 & -0.50 \\ 
$\Lambda(2350)$ & 8 & 1 &2350 & 2335.67 & -0.61 \\ 
$\Xi^{0}_c(2471)$ & 8 & 2 &2471 & 2472.47 & 0.06 \\ 
$\Lambda_c^+(2593)$ & 8 & 3 &2595 & 2593.89 & -0.06 \\ 
$\Omega_c^0$ & 8 & 4 &2698 & 2703.03 & 0.21 \\ 
$\Sigma_c^0(2800)$ & 8 & 5 &2800 & 2802.16 & 0.08 \\ 
$\Lambda_c^+(2880)$(***) & 8 & 6 &2882 & 2892.98 & 0.38 \\ 
$\Xi(2980) $(***) & 8 & 7 &2978 & 2976.79 & -0.06 \\ 
$\Xi_c(3080)$(***) & 8 & 8 &3076 & 3054.61 & -0.70 \\ 
$\Omega^-(2380)$(**) & 9 & 0 &2380 & 2379.27 & -0.03 \\ 
$\Sigma_c(2520)$(***) & 9 & 1 &2518 & 2535.82 & 0.69 \\ 
$\Sigma_c(2645)$(***) & 9 & 2 &2646 & 2672.62 & 1.00 \\ 
\hline\noalign{\smallskip}
$\Lambda_c(2880)$(***) & 9 & 4 &2882 & 2903.18 & 0.74 \\ 
$\Sigma(3170)(*)$ & 9 & 7 &3170 & 3176.94 & 0.22 \\ 
$\Xi_{cc}^+(*)$ & 10 & 9 &3519 & 3517.06 & -0.05 \\ 
$\Omega_{cc}$ ({\cite{th}})& 11 & 8 &3637 & 3627.63 &  \\ 
$\Omega_{ccc}$({\cite{th}}) & 15 & 15 &4681 & 4702.20 &  \\ 
$\Lambda_b^0$(***) & 20 & 20 &5620 & 5612.75 & -0.13 \\ 
\end{tabular} 
\end{center}
\end{table} 

The derived equation  is a fractional integro-differential equation, a solution may be obtained by iteration. 
In lowest order approximation, we set 
\begin{equation} 
\label{approx}
z^{\alpha-1} \hat{L}_z(2\alpha-1) \approx \rho \hat{L}_z(\alpha) 
\end{equation}
with the constant $\rho$ and obtain
\begin{equation}
\label{zeeman0}
H^0 \psi^0 =  - {1 \over 2 m^{2 \alpha-1}}  
(\Delta^\alpha_\alpha  +  \bar{g}\alpha\Gamma(2-\alpha) B \rho \hat{L}_z(\alpha)  )\psi = E^0 \psi^0
\end{equation}

Since $\hat{L}_z(\alpha)$ is the Casimir operator of the fractional rotation group $SO^\alpha(2)$, for
an analytic solution 
a group theoretical approach is appropriate \cite{he07}.

In order to classify the multiplets only, the 
 Hamiltonian (\ref{zeeman0}) may be rewritten as a linear combination of the Casimir operators
$\hat{L}^2(\alpha)$ of $SO^\alpha(3)$ which corresponds in a classical picture to the fractional 
angular momentum  and $\hat{L}_z(\alpha)$ of $SO^\alpha(2)$ which corresponds to the z-projection of the 
angular momentum.  The eigenfunctions are determined by two
quantum numbers $\mid \! LM\!>$
\begin{equation}
\label{su}
 H^0 \mid \! LM\!>= m_0 + a_0 \hat{L}^2 + b_0 \hat{L}_z \mid \! LM\!> 
\end{equation}
The eigenvalues of the Casimir operators based on the Riemann-Liouville 
fractional derivative definition  are given by \cite{he07}
\begin{eqnarray}
\label{eqLz}
\hat{L}_z(\alpha) \mid\! LM\!> & = & \pm \frac{\Gamma(1+ \mid\!\! M \!\!\mid\! \alpha)}{\Gamma(1+( \mid\!\! M \!\!\mid\! -1)\alpha)}  \mid  \!LM\!>  \nonumber \\
                & &  \qquad  M=\pm 0,\pm 1,\pm 2,...,\pm L  \\
\label{eqL2}
\hat{L}^2(\alpha) \mid\! LM\!> & = &  \frac{\Gamma(1+ (L+1) \alpha)}{\Gamma(1+(L-1)\alpha)}  \mid \!LM\!> \nonumber \\ 
             & & \qquad   L=0,+1,+2,... 
\end{eqnarray}
where $\mid\!\! M \!\!\mid$ is the absolute value of $M$.
Note that for $\alpha=1$ the Casimir operators and the corresponding eigenvalues reduce to the
well known results of 
 standard quantum mechanical angular momentum algebra \cite{edmonds}.  

With (\ref{eqLz}) and (\ref{eqL2}) the level spectrum of the multiplets  is  determined
\begin{equation}
\label{zeemanlevels}
E^0 = m_0 + 
a_0 {\Gamma(1 + (L+1)\alpha) \over \Gamma(1 + (L-1) \alpha)}  \nonumber \\
          \pm b_0 \frac{\Gamma(1+ \mid\!\! M \!\!\mid\! \alpha)}{\Gamma(1+( \mid\!\! M \!\!\mid\! -1)\alpha)}   
\end{equation}
\vspace{-7mm}
\subsection{Phenomenology of the baryon spectrum}
We have derived in lowest order an analytic expression (\ref{zeemanlevels}) for 
the splitting of the energy levels of a  non relativistic charged fractional  spinless particle 
in  a constant fractional magnetic field. If we associate this particle with a quark and the fractional magnetic
field with a color magnetic field, this model
should allow  a description of the hadron spectrum. 

This assumption is motivated by results 
obtained by ab initio calculations in lattice QCD \cite{wilson}, where a color flux tube with almost constant 
color magnetic field between quarks is observed. 

Since we expect a similar result for self consistent solutions of 
the nonlinear field equations (\ref{fdirac}) and (\ref{fmaxwell}), the fractional Zeeman-effect serves as 
a simple, idealized model for a first test, whether  the hadron spectrum  may be reproduced at all
within the framework of a fractional abelian field theory.

Therefore we will use formula (\ref{zeemanlevels})  for a fit of the experimental baryon spectrum \cite{pdg}. 

The optimum fit parameter set with an r.m.s error of $0.84\%$ results as:
\begin{eqnarray}
\alpha  &=&   0.112\\
m_0 &=&  -17171.6 [MeV] \\
a_0 &=& 10971.8 [MeV]\\
b_0 &=& 8064.6[MeV] 
\end{eqnarray}
In figure \ref{had} the experimental versus fitted values are shown. 
In table \ref{tone} name, proposed quantum numbers $L$ and $M$, experimental mass $E^{exp}[MeV]$ 
taken from \cite{pdg}, fitted
mass $E^{th}[MeV]$ according to (\ref{zeemanlevels}) and  error  $\Delta$E[$\%$] are listed. 

Remarkable enough, the overall error for the fitted baryon spectrum $L=3,..,9$ is less than 1 $\%$. A general trend
is an increasing error for small masses which possibly
indicates the limitations of a non relativistic model.

The $\alpha=0.112$ value corresponds in the sense of the relativistic theory (see (\ref{alpha}))
to an inherent  $SU(18)$ symmetry,  which
optimistically may be interpreted as the direct product:  
\begin{equation}
SU(18) = SU(N=6)^{flavour} \otimes SU(3)^{color}
\end{equation}
and is therefore an indirect  indication for the maximum number of different quark flavours to be found in nature.

In the lower part of table \ref{tone} the proposed $(L,M)$ values for $\Lambda_c(2880)$ up to $\Lambda_{b}$
give a rough estimate for the number of particles,  which are still missing in the experimental spectrum.

For $L=1,2$ we can compare the theoretical mass predictions for baryons
with the experimental meson spectrum, see upper part of table \ref{tone} and left part of figure \ref{had}. 
The results are
surprisingly close to the experimental masses, despite the fact that mesons and baryons are different constructs.
This means, that the relative strength of the fractional magnetic field $b_0 = B_q/m_q$, $\{ q \in u,d,s,c   \}$ is of 
similar magnitude in mesonic ($q\bar{q}$) and baryonic $qqq$ systems. 

Summarizing the results of this section we conclude, that the fractional Zeeman effect, which describes
the motion of a fractional charged particle in a constant fractional magnetic field serves as a reasonable
non relativistic model for an understanding of the full baryon spectrum.

This result has far reaching consequences:
For the first time the terms "particle described by a fractional wave equation"  and 
"fractional magnetic field"     have successfully been
associated with the entities quark, which follows a $SU(n)^{flavour}$ symmetry and color, which follows a
$SU(3)^{color}$ symmetry.

Besides classical QED a set of wave equations 
e.g. the Dirac-, Pauli- and Schr\"odinger equation
describe the interaction of electrons, positrons and photons with descending accuracy. The results derived
in this section indicate, that this probably also holds for fractional QED.

Furthermore, the results indicate, that  a fractional QED based on the Lagrange density (\ref{ldirac}) with a small coupling constant $\bar{g}$ 
and an abelian gauge field
and QCD based on the standard QCD Lagrange density with a strong coupling 
constant $\bar{g}$ 
and a non abelian gauge field
may probably  describe the same phenomena. 

The results presented so far, encourage further studies in this field. The next step is a thorough investigation of
the transformation properties of the proposed Lagrangians under fundamental symmetry transformations. 
In the relativistic case, 
invariance of the Lagrangian and corresponding wave equations under Lorentz-transformations has been investigated for
free fields e.g. in \cite{zavada}. 

For example in case of the ordinary Dirac equation the transformation laws 
for its solution and simultaneously for the operator itself are automatically 
the same for free equation and for the equation with the usual interaction. On 
the other hand in \cite{zavada} proof of invariance is given only for free fractional 
fields and it is obvious that invariance of corresponding fields with 
interaction represents a further and probably much more complicated problem. And 
without existence of transformation laws for all fields it is not possible to 
assume that related Lagrangian is a scalar quantity.

In the non relativistic case, invariance under spatial rotations is a transformation
of similar importance. Therefore in the following section we will investigate the behaviour of the fractional
Schr\"odinger equation under spatial rotations.    

\section{The internal structure of fractional particles}
Properties of particles, which are described by wave equations, may be investigated using the
commutation relations of fundamental symmetry operations.

Lets call a particle elementary, if it is described by a potential- and field-free wave equation. If in addition
there is an internal structure, which is determined by additional quantum numbers, it may be revealed
e.g. considering the behaviour under rotations.

We will investigate the most simple case, the behaviour of the fractional Schr\"odinger equation (\ref{sgl})
under rotations in $R^2$ about the z-axis.

Using the Leibniz product rule (\ref{leibniz}) the fractional derivative of the
product $x f(x)$ is given by 
\begin{eqnarray}
\label{leibnizxf}
\partial^\alpha_x (x f(x)) &=& \sum_{j=0}^\infty    
\left(\begin{array}{c}
\alpha \\ j \\
\end{array} \right)
(\partial^j_x x) \partial^{\alpha-j}_x f(x) \\
&=& \left(x \partial^\alpha_x + \alpha \partial^{\alpha-1}_x \right) f(x)
\end{eqnarray}
We define a generalized fractional angular momentum operator 
$K^\beta$ with z-component $K^\beta_z$
\begin{equation}
\label{kz}
K^\beta_z = i \left({\hbar \over mc}\right)^\beta \, mc\,  (x \partial^\beta_y - y \partial^\beta_x) 
\end{equation}
The components $K^\beta_x, K^\beta_y$ are given by cyclic permutation of the spatial indices in (\ref{kz}). 
Using (\ref{leibnizxf})
the commutation relation  with the Hamiltonian $H^\alpha$ of the free Schr\"odinger equation (\ref{sgl}) results as
(with units $m=1$, $\hbar=1$, $c=1$):
\begin{equation}
\label{kkk}
[ i  K^\beta_z, H^\alpha] =      \alpha \left(    
                          \partial^{2 \alpha-1}_x   \partial^\beta_y 
                         - \partial^\beta_x \partial^{2 \alpha-1}_y 
                          \right) 
 \end{equation}
Setting $\beta=1$ we obtain for the z-component of the standard angular momentum operator  $L_z$ the 
commutation relation
\begin{equation}
[L_z,H^\alpha] = [K^{\beta=1}_z,H^\alpha] =  -  i \alpha \left(    
                          \partial^{2 \alpha-1}_x   \partial_y 
                         - \partial_x \partial^{2 \alpha-1}_y 
                          \right) 
\end{equation}
which obviously is not vanishing. Therefore particles described with the 
fractional Schr\"o\-dinger equation (\ref{sfree}) contain an internal structure for $\alpha \neq 1$.

We now define the fractional total angular momentum 
$J^\beta$ 
with z-component $J^\beta_z$.
Setting 
\begin{equation}
\beta = 2 \alpha-1
\end{equation}
we obtain with $J^{2 \alpha-1}_z = K^{2 \alpha-1}_z$, 
and with (\ref{kkk}) 
\begin{equation}
[J^{2 \alpha-1}_z,H^\alpha ] = 0
\end{equation}
this operator commutes with the Hamilton operator. Therefore $J^{2 \alpha-1}_z$ indeed is the
fractional analogue of the standard z-component of the total angular momentum.

Now we define the z-component of a fractional intrinsic angular momentum $S^\beta_x$
with
\begin{equation}
J^{2 \alpha-1}_z = L_z + S^{2 \alpha-1}_z 
\end{equation}
The explicit form is given by
\begin{eqnarray}
S^{2 \alpha-1}_z &=&  i 
 x \left(\left({\hbar \over mc}\right)^{2 \alpha-1} \! mc\, \partial^{2 \alpha-1}_y - \hbar \partial_y\right)  \\ \nonumber
& &- i y \left(\left({\hbar \over mc}\right)^{2 \alpha-1} \! mc\,\partial^{2 \alpha-1}_x - \hbar \partial_x\right)   
\end{eqnarray}
This operator vanishes for $\alpha=1$, whereas for $\alpha \neq 1$ it gives the z-component of a 
fractional spin.

Lets call the difference between fractional and ordinary derivative  $\delta p$, or more precisely
the components
\begin{equation}
\delta p_i = i  \left(\left({\hbar \over mc}\right)^{2 \alpha-1} mc\,\partial^{2 \alpha-1}_i - \hbar \partial_i \right) 
\end{equation}
for $S^{2 \alpha-1}_z$ we can write
\begin{equation}
S^{2 \alpha-1}_z =   x \delta p_y - y \delta p_x 
\end{equation}
The components of a fractional spin vector are then given by the cross product
\begin{equation}
\vec{S}^{2 \alpha-1} =   \vec{r} \times \delta \vec{p}
\end{equation}
Therefore fractional spin describes an internal fractional rotation, which is proportional to the 
momentum difference between fractional and ordinary  momentum. For a given $\alpha$ it has exactly
one component. 

With $J^{2 \alpha-1}_x= K^{2 \alpha-1}_x$ and $J^{2 \alpha-1}_y=K^{2 \alpha-1}_y$, 
the commutation relations for the total fractional angular momentum are given by
\begin{eqnarray}
\left[ J^{2 \alpha-1}_x, J^{2 \alpha-1}_y \right] &=&  (2 \alpha-1) {\hbar \over m c} J^{2 \alpha-1}_z p_z^{2(\alpha-1)} \\
\left[ J^{2 \alpha-1}_y, J^{2 \alpha-1}_z \right] &=&  (2 \alpha-1) {\hbar \over m c} J^{2 \alpha-1}_x p_x^{2(\alpha-1)} \\
\left[ J^{2 \alpha-1}_z, J^{2 \alpha-1}_x \right] &=&  (2 \alpha-1) {\hbar \over m c} J^{2 \alpha-1}_y p_y^{2(\alpha-1)} 
\end{eqnarray}
with components of the momentum operator $p$ or generators of fractional translations respectively given as
\begin{equation}
p_i^\beta =   i \left( {\hbar \over m c} \right)^\beta m c \, \partial^\beta_i 
\end{equation}
Therefore in the general case $\alpha \neq 1$ an extended fractional rotation group is generated, which contains 
an additional fractional translation factor.

Hence we conclude, that fractional elementary particles carry an internal structure, which we call
fractional spin, because analogies to e.g. electron spin are close. 

Consequently, the transformation properties of the fractional Schr\"o\-dinger equation 
are more related to the ordinary Pauli-equation
than to the ordinary ($\alpha=1$) Schr\"o\-dinger-equation.  

In addition, the results presented lead to the conclusion, that the fractional magnetic field $B^\alpha$ interacts
with the fractional angular momentum and the intrinsic fractional angular momentum simultaneously in a similar manner as
in the case with particles, described with the Pauli-equation. 

If we introduce a fractional external field $\vec{A}$ of the type:
\begin{equation}
\label{aconst2}
\vec{A}  = \{ - {1 \over 2} B x^{1-\alpha} y z^{\alpha-1},{1 \over 2} B x y^{1-\alpha}z^{\alpha-1},0 \}  
\end{equation}
which for $\alpha=1$  reduces to $ \{-\frac{1}{2}B y,\frac{1}{2}B x ,0 \} $ like (\ref{aconst}),we obtain the fractional 
Schr\"odinger-equation in analogy to (\ref{zeeman}) as:
\begin{equation}
\label{zeeman2}
H \psi =  - {1 \over 2 m^{2 \alpha-1}}  
(\Delta^\alpha_\alpha  +  \bar{g}\alpha\Gamma(2-\alpha) B z^{\alpha-1} J_z^{2\alpha-1}  )\psi = E \psi
\end{equation}
which clearly demonstrates the close relationship with the Pauli equation.

\section{Conclusion}
The principle of local gauge invariance has been applied to fractional fields. As a result, we obtained
the exact form of an interaction up to order $o(\bar{g})$ which corresponds to an invariance 
under infinitesimal gauge transformations.

 The resulting non linear field equations
are extensions of the QED equations, where 
 the electric field $A_\mu$ extends from a c-number to a non local operator, which we 
called the fractional charge     connection operator and where the 
$\gamma_\alpha^\mu$-matrices obey an extended Clifford-algebra, which reflects the transition from $SU(2)$ to $SU(n>2)$.

This makes the fractional QED an interesting candidate to describe properties of particles and gauge fields, 
which obey
an inherent $SU(n)$ symmetry.

As a first application, we derived an analytic mass formula of a non relativistic fractional
 charged particle moving in a
constant fractional magnetic field. 

Associating the particle with a quark and the fractional magnetic field with the color 
magnetic field,
this formula reproduces the full baryon spectrum with an error less than $1\%$.      

Therefore the step from free  to interacting fractional fields based on the principle of local 
gauge invariance opens up 
a new exciting research area in fractional calculus.    



\begin{thebibliography}{00}
%
\bibitem{pauli} Pauli W   Rev.Mod.Phys. {\bf 13}, (1941) 203.
\bibitem{sakurai}  Sakurai J J  1967  {\it Advanced Quantum Mechanics} Benjamin/Cummings, New York.
\bibitem{yang} Yang C N and Mills R L Phys. Rev. {\bf 96}, (1954) 191.
\bibitem{lee} Abers E S and Lee B W  Physics Reports  {\bf C9}, (1973) 1.
\bibitem{f1} Leibniz  G F Sep 30, 1695 {\it Correspondence with l`Hospital}, manuscript.
\bibitem{f2}Liouville J 1832 J. $\acute{E}$cole Polytech.,  {\bf 13}, 1-162.
\bibitem{riemann}  Riemann B Jan 14, 1847 {\it  Versuch einer allgemeinen Auffassung der Integration
und Differentiation } in:  Weber H (Ed.), {\it Bernhard Riemann's gesammelte mathematische Werke und wissenschaftlicher
Nachlass},  Dover Publications (1953),  353.
\bibitem{raspini} Raspini A  Fizika B {\bf 9}, (2000) 49.
\bibitem{raspini1} Raspini A  Physica Scripta {\bf 64}, (2001) 20.
\bibitem{zavada} Zavada P  J Appl. Math. {\bf 2:4}, (2002) 163.
\bibitem{lim} Lim C and Muniandy S V   Phys.Lett.A {\bf 324}, (2004) 396.
\bibitem{gold}  Goldfain E  Chaos, Solitons and Fractals {\bf 28}, (2006)  913
\bibitem{zeeman} Zeeman P  Nature {\bf 55}, (1897) 347.
\bibitem{f3}  Miller K  and  Ross B 1993  {\it An Introduction to Fractional Calculus and Fractional Differential
Equations} Wiley, New York.
\bibitem{o1} Oldham K B and Spanier J 2006 {\it The Fractional Calculus},
Dover Publications, Mineola, New York.
\bibitem{he08}  Herrmann R 2008 {\it Fraktionale Infinitesimalrechnung - Eine Einf\"uhrung f\"ur Physiker}, BoD, Norderstedt, Germany
\bibitem{caputo} Caputo M  Geophys. J. R. Astr. Soc.  \textbf{13}, (1967) 529.
\bibitem{weyl} Weyl H  Vierteljahresschrift der Naturforschenden
Gesellschaft in Z\"urich  \textbf{62}, (1917) 296.
\bibitem{riesz} Riesz M  Acta Math.   \textbf{81}, (1949) 1.
\bibitem{feller} Feller W Comm. Sem. Mathem. Universite de Lund, (1952) 73-81.
\bibitem{grun} Gr\"unwald A K Z. angew. Math. und Physik   \textbf{12}, (1867) 441.
\bibitem{pod} Podlubny I 1999 {\it Fractional Differential equations},
Academic Press, New York.
\bibitem{he07}  Herrmann R J. Phys. G: Nucl. Part. Phys.  {\bf 34}, (2007), 607
\bibitem{baleanu} Baleanu D and Muslih S  Physica Scripta \textbf{72}, (2005) 119.
\bibitem{baleanu1} Baleanu D and Tas K math-ph/0612024
\bibitem{edmonds} Edmonds A R  1957   {\it Angular Momentum in Quantum Mechanics}, 
Princeton University Press,  New Jersey.
\bibitem{wilson} Wilson K G  Phys. Rev. D \textbf{10} (1974) 2445
\bibitem{pdg} Yao W M et al. (particle data group) J. Phys. G  {\bf 33}, (2006) 1. 
and 2007 updates for the 2008 edition
\bibitem{th} Chiu T. and Hsieh T. hep-lat/0501021, (2005) 
\end{thebibliography}
\end{document}